\documentclass[aps,prl,showpacs,superscriptaddress,floatfix,10pt]{revtex4}  

\usepackage{graphicx}

\newcommand{\beq}[1]{\begin{equation}\label{#1}}
\newcommand{\eeq}{\end{equation}}
\newcommand{\bea}[1]{\begin{eqnarray}\label{#1}}
\newcommand{\eea}{\end{eqnarray}}

\newcommand{\eq}[1]{(\ref{#1})}
\newcommand{\Eq}[1]{Eq.~(\ref{#1})}

\newcommand{\Fi}[1]{Fig.~\ref{#1}}
\newcommand{\Ta}[1]{Table~\ref{#1}}

\newcommand{\agev}{\mbox{$A$~GeV}}               

\newcommand{\rb}[1]{\mbox{\textrm{\scriptsize #1}}}

\newcommand{\ab}[1]{\langle #1 \rangle}

\newcommand{\pimin}{\ensuremath{\pi^-}}
\newcommand{\piplus}{\ensuremath{\pi^+}}
\newcommand{\piplusmin}{\ensuremath{\pi^{\pm}}}
\newcommand{\kmin}{\ensuremath{\textrm{K}^-}}
\newcommand{\kplus}{\ensuremath{\textrm{K}^+}}
\newcommand{\kplusmin}{\ensuremath{\textrm{K}^{\pm}}}
\newcommand{\kbar}{\ensuremath{\overline{\textrm{K}}}}
\newcommand{\neutron}{\ensuremath{\textrm{n}}}
\newcommand{\proton}{\ensuremath{\textrm{p}}}
\newcommand{\pbar}{\ensuremath{\overline{\textrm{p}}}}
\newcommand{\eplus}{\ensuremath{\textrm{e}^+}}

\newcommand{\pT}{\ensuremath{p_{\rb{T}}}}

\newcommand{\mT}{\ensuremath{m_{\rb{T}}}}
\newcommand{\dedx}{\ensuremath{\textrm{d}E/\textrm{d}x}}
\newcommand{\dndy}{\ensuremath{\textrm{d}n/\textrm{d}y}}
\newcommand{\enwee}{\ensuremath{N_{\rb{w}}}}
\newcommand{\nwound}{\ensuremath{\langle \enwee \rangle}}
\newcommand{\eejes}{\ensuremath{E_{\rm S}}}

\newcommand{\Tc}{\ensuremath{T_{\rb{c}}}}

\newcommand{\der}{\ensuremath{\textrm{d}}}

\newcommand{\tofdedx}{\mbox{TOF + \dedx}}

\begin{document}




\title{Pion and kaon production in central Pb+Pb collisions at 20$A$ and
30$A$ GeV: \\
Evidence for the onset of deconfinement 
}





\affiliation{NIKHEF, Amsterdam, Netherlands.}  
\affiliation{Department of Physics, University of Athens, Athens, Greece.}
\affiliation{Comenius University, Bratislava, Slovakia.}
\affiliation{KFKI Research Institute for Particle and Nuclear Physics,
             Budapest, Hungary.}
\affiliation{MIT, Cambridge, USA.}
\affiliation{Henryk Niewodniczanski Institute of Nuclear Physics,
             Polish Academy of Science, Cracow, Poland.}
\affiliation{Gesellschaft f\"{u}r Schwerionenforschung (GSI),
             Darmstadt, Germany.} 
\affiliation{Joint Institute for Nuclear Research, Dubna, Russia.}
\affiliation{Fachbereich Physik der Universit\"{a}t, Frankfurt, Germany.}
\affiliation{CERN, Geneva, Switzerland.}
\affiliation{Institute of Physics \'Swi{\,e}tokrzyska Academy, Kielce, Poland.}
\affiliation{Fachbereich Physik der Universit\"{a}t, Marburg, Germany.}
\affiliation{Max-Planck-Institut f\"{u}r Physik, Munich, Germany.}
\affiliation{Institute of Particle and Nuclear Physics, Charles
             University, Prague, Czech Republic.}
\affiliation{Department of Physics, Pusan National University, Pusan,
             Republic of Korea.} 
\affiliation{Nuclear Physics Laboratory, University of Washington,
             Seattle, WA, USA.} 
\affiliation{Atomic Physics Department, Sofia University St.~Kliment
             Ohridski, Sofia, Bulgaria.} 
\affiliation{Institute for Nuclear Research and Nuclear Energy, Sofia, Bulgaria.}
\affiliation{Department of Chemistry, Stony Brook University (SUNYSB), Stony Brook, USA.}
\affiliation{Institute for Nuclear Studies, Warsaw, Poland.}
\affiliation{Institute for Experimental Physics, University of Warsaw,
             Warsaw, Poland.} 
\affiliation{Faculty of Physics, Warsaw Univeristy of Technology, Warsaw, Poland.}
\affiliation{Rudjer Boskovic Institute, Zagreb, Croatia.}


\author{C.~Alt}
\affiliation{Fachbereich Physik der Universit\"{a}t, Frankfurt, Germany.}
\author{T.~Anticic} 
\affiliation{Rudjer Boskovic Institute, Zagreb, Croatia.}
\author{B.~Baatar}
\affiliation{Joint Institute for Nuclear Research, Dubna, Russia.}
\author{D.~Barna}
\affiliation{KFKI Research Institute for Particle and Nuclear Physics,
             Budapest, Hungary.} 
\author{J.~Bartke}
\affiliation{Henryk Niewodniczanski Institute of Nuclear Physics,
             Polish Academy of Science, Cracow, Poland.}
\author{L.~Betev}
\affiliation{CERN, Geneva, Switzerland.}
\author{H.~Bia{\l}\-kowska} 
\affiliation{Institute for Nuclear Studies, Warsaw, Poland.}
\author{C.~Blume}
\affiliation{Fachbereich Physik der Universit\"{a}t, Frankfurt, Germany.}
\author{B.~Boimska}
\affiliation{Institute for Nuclear Studies, Warsaw, Poland.}
\author{M.~Botje}
\affiliation{NIKHEF, Amsterdam, Netherlands.}
% ...


\collaboration{The NA49 collaboration}
\noaffiliation


\begin{abstract}

Results on charged pion and kaon production in central Pb+Pb
collisions at 20$A$ and 30\agev\ are presented and compared
to data at lower and higher energies.  A rapid change of the
energy dependence is observed around 30\agev\ for the yields of pions
and kaons as well as for the shape of the transverse mass spectra.
The change is compatible with the prediction that the threshold for
production of a state of deconfined matter at the early stage of 
the collisions is located at low SPS energies.

\end{abstract}

\pacs{...}




\maketitle

\section{Introduction}

The advent of the quark model of hadrons and the development of
quantum chromodynamics naturally led to the question whether strongly
interacting matter exists in different phases and, if so, of which
nature the transitions between these phases are.  In particular, it is
commonly believed that a gas of hadrons will undergo a transition to a
state of quasi-free quarks and gluons, the Quark Gluon Plasma
(QGP)~\cite{qgp}, when its temperature exceeds a critical
value~\cite{lqcd}.  These questions have motivated a broad
experimental program of nucleus-nucleus collisions to study the
properties of strongly interacting matter at extreme densities and
temperatures.

Several signatures of the formation of a transient QGP state during
the early stage of a nucleus-nucleus collision at high energies were
proposed in the past~\cite{Ra:82}. However, the validity of these
signatures has come under renewed scrutiny. For this reason the NA49
Collaboration at the CERN SPS has searched over the past years for
signs of the onset of QGP creation in the energy dependence of hadron
production.  This search was motivated by the prediction
of~\cite{GaRo,Ga:95,GaGo} that the onset of deconfinement should lead
to a steepening of the increase of the pion yield with collision
energy and to a sharp maximum in the energy dependence of the
strangeness to pion ratio.  The onset was expected
to occur at approximately 30\agev~\cite{GaGo}.

The NA49 energy scan program at the CERN SPS started with two runs
where data on central Pb+Pb collisions at 40$A$ and 80\agev\ were
recorded in 1999 and 2000. Data at the top SPS energy of 158\agev\ had
already been taken in previous SPS runs.  The analysis of these runs
was published in~\cite{Afanasiev:2002mx} and the results confirmed the
predictions of~\cite{GaGo}. This finding motivated an extension of the
energy scan to the lower SPS energies of 30$A$ and 20\agev\ which was
completed in 2002.  In this letter we report final results on charged
kaon and pion production in central Pb+Pb collisions at 20$A$ and 
30\agev\ beam energy.  The present measurements are combined with those
from~\cite{Afanasiev:2002mx} and compared to available model
calculations, together with lower and higher energy data from AGS and
RHIC.


\section{Detector and data analysis}

The NA49 experimental set-up~\cite{na49_nim} consists of four large
volume Time Projection Chambers (TPC).  Two of these (VTPC) are placed
in the field of two super-conducting dipole magnets.  The other two
TPCs (MTPC) are positioned downstream of the magnets and are optimized
for high precision measurements of the ionization energy loss \dedx\ 
with a resolution of about 4\%. This \dedx\ measurement provides
particle identification which is complemented by a measurement of the
time-of-flight (TOF) with a resolution of about 60~ps in two TOF
detector arrays positioned downstream of the MTPCs.  At each incident
energy the TOF acceptance for kaons was kept at mid-rapidity by
lowering in proportion to the beam energy the nominal 158\agev\ field
settings of about 1.5 (upstream magnet) and 1.1~T (downstream magnet).
In both the 30$A$ and 20\agev\ runs a thin lead foil target of 224~mg/cm$^2$,
corresponding to about 1\% of interaction length for Pb ions, was
positioned 80~cm upstream from the first TPC.  A trigger based on a
measurement of the energy deposit of projectile spectator nucleons in
a downstream calorimeter selected the most central 7.2\% of the Pb+Pb
collisions.  The corresponding mean number of wounded
nucleons $\enwee$~\cite{bialas} is calculated using the Fritiof model~\cite{Fr}
to be $\enwee = 349 \pm 1 (\mbox{stat}) \pm5 (\mbox{sys})$.  At each
energy about $3.5 \times 10^5$ events were recorded.

Results on kaon and pion production were obtained in a multi-step
analysis procedure which involves charged track reconstruction,
particle identification and corrections to account for background
contributions as well as for acceptance and efficiency losses.  To reduce
the systematic errors, the analysis has been restricted to regions of
phase space where background and efficiency corrections are small and
approximately uniform.  To minimize tracking efficiency
corrections, only tracks within an azimuthal angle wedge of
$\pm30^o$ with respect to the horizontal plane were used.

The spectra of $\pi^\pm$ and K$^\pm$ at mid-rapidity are obtained
using the combined \dedx\ and TOF information as described
in~\cite{na49_ppbar} (\tofdedx\ analysis).  The pion spectra were
corrected for the contribution from the weak decays of the strange
particles $\Lambda$, $\Sigma^0$, $\Sigma^{\pm}$ and $K^0_S$ as well as
for $\mu$ contamination (charged products of weak decays can
  be reconstructed as coming from the event interaction vertex, while
  muons cannot be separated from pions in NA49 with \dedx\ and TOF
  measurements for momenta above about 1~GeV/c).  The correction
factors were obtained from a GEANT simulation of the NA49 detector
using as input the hadron distributions from the VENUS
model~\cite{venus}.  The model distributions were tuned to reproduce
the measured yields of $\Lambda$~\cite{na49_lambda20-30} and $K^0_S$
where the latter were taken to be the average of the \kplusmin\ yields
presented in this paper. The total background correction to the
$\pi^-$ ($\pi^+$) yields is found to be about 8 (6)\%.

Raw \kplus\ and \kmin\ yields at forward rapidities were extracted
from fits of the \dedx\ distributions in narrow bins of momentum and
transverse momentum.  The fitted function parametrizes contributions
from \eplus, $\pi^+$, \kplus, protons and deuterons to the \dedx\ 
spectra of positively charged particles and corresponding
contributions to those of negatively charged particles.
For pions the \dedx\ method does not provide sufficiently accurate 
identification in the low \pT\ region near mid-rapidity.
Therefore to obtain the raw $\pi^-$ spectra in the full forward hemisphere
of the reactions yields
of all negatively charged particles were determined as a function of
rapidity (calculated assuming the pion mass) and transverse momentum
\pT.  The contamination by \kmin, \pbar\ and e$^-$ from
the interaction vertex as well as non--vertex hadrons originating from
strange particle decays and secondary interactions was subtracted.
The total correction amounts to 20-25\% and was calculated using
the VENUS/GEANT simulation. The above procedure can not be applied for
positively charged hadrons due to the large contribution of kaons and
protons.

The resulting background subtracted spectra were corrected for
geometrical acceptance, losses due to in-flight decays,
inefficiencies of the tracking algorithms (about 5\%) and quality
cuts.  More details on the correction procedure can be found
in~\cite{Afanasiev:2002mx}.

Systematic errors were estimated by comparing results obtained with
different detectors (TPC, TOF) and by varying cuts and correction
strategies.  Uncertainties in the parameters of the \dedx\ fits lead
to asymmetric systematic errors on the kaon yields.  The systematic
errors were estimated to be 5-10\% and are explicitly given in
Table~II. Note that to a large extent the systematic uncertainties
presented here are common to those reported in~\cite{Afanasiev:2002mx}
since there the same experimental procedure was used.

\section{Results at 20$A$ and 30$A$ GeV}

In \Fi{fig1}
%
\begin{figure}[htb]
\mbox{ \includegraphics[width=0.95\linewidth]{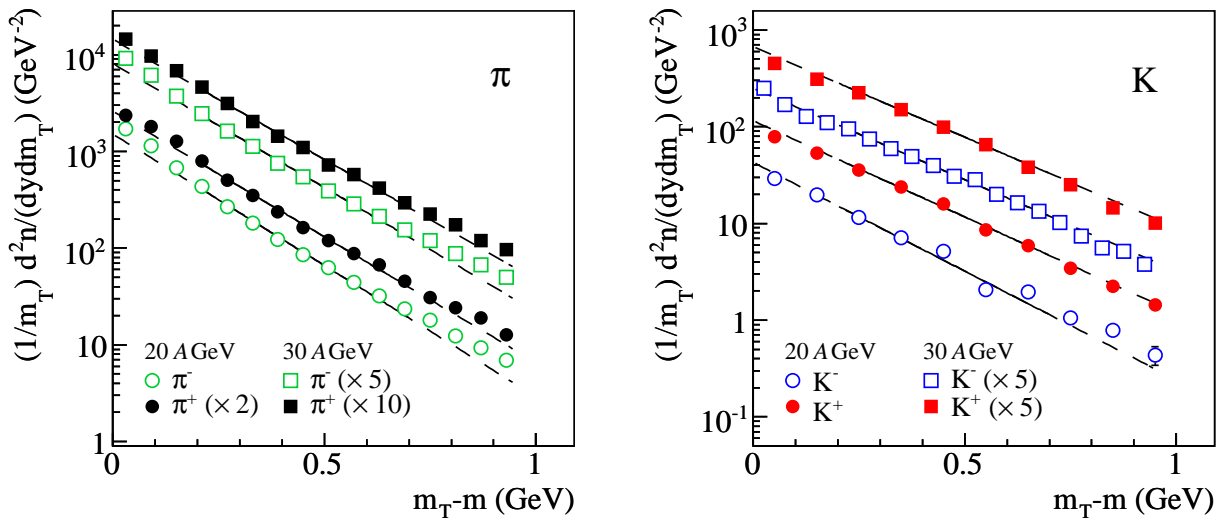}}
\caption{\label{fig1} 
  Transverse mass spectra of $\pi^-$ and $\pi^+$ mesons at $0 < y <
  0.2$ (left) and those of K$^+$ and K$^-$ mesons at $|y| < 0.1$
  (right) produced in central Pb+Pb collisions at 20$A$ and 30\agev.
  The lines are fits of \Eq{mt} to the spectra in the interval $0.2 <
  \mT -m  < 0.7$~GeV.  The statistical errors are smaller than the
  symbol size.  The systematic errors are 5\% in the region used for
  the fit and reach 10\% at the low and high ends of the \mT\ spectra.
}
\end{figure}
%
are shown the transverse mass spectra of $\pi^{\pm}$ mesons near
mid-rapidity $0 < y < 0.2$ together with those of \kplusmin\ at $|y| <
0.1$ obtained from the \tofdedx\ analysis of 20$A$ and 30\agev\ central
Pb+Pb collisions.  Here the transverse mass is defined by $\mT^2 =
\pT^2 + m^2$ with $m$ the rest mass of the particle, and $y$ denotes
the rapidity of a particle in the collision center-of-mass system.
The full lines in \Fi{fig1} indicate a fit of the function
\beq{mt}
  \frac {\der^2 n} {\mT \der \mT \der y} = C \, \exp \left( - \frac
  {\mT} {T} \right) 
\eeq
to the data in the range $0.2 < \mT -m < 0.7$~GeV.  The values
obtained for the inverse slope parameter $T$ are given in \Ta{tab:results3}.
%

The kaon spectra are well described by the fit, while the low and high
\mT\ regions of the pion spectra outside of the fitted region are
above the extrapolated fitted line.

The rapidity distributions $\der n / \der y$ of $\pi^-$, \kmin\ and \kplus\ 
mesons are plotted in \Fi{fig2}.
%
\begin{figure}[htb]
\mbox{ 
       \includegraphics[width=0.33\linewidth]{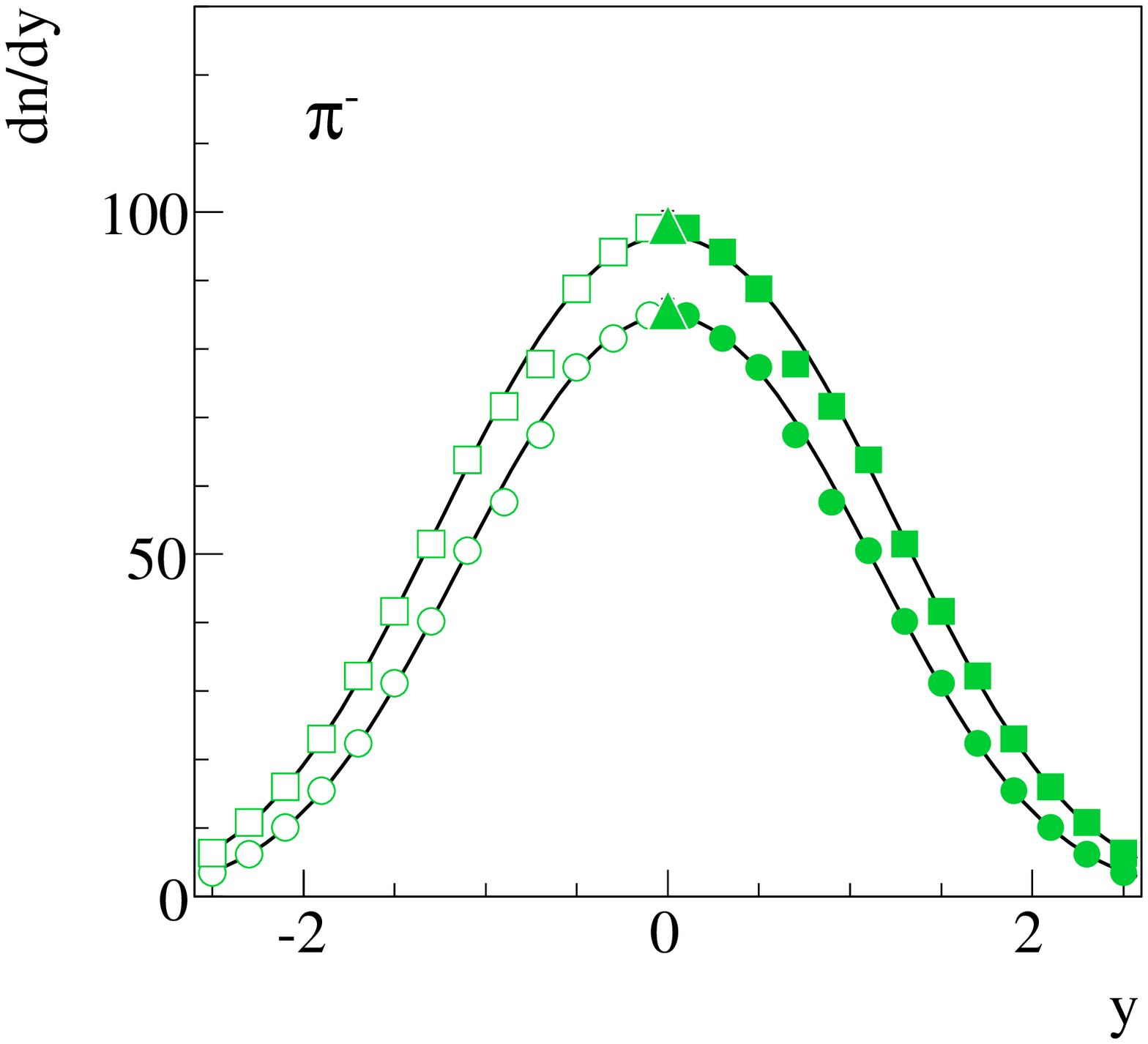}
       \includegraphics[width=0.33\linewidth]{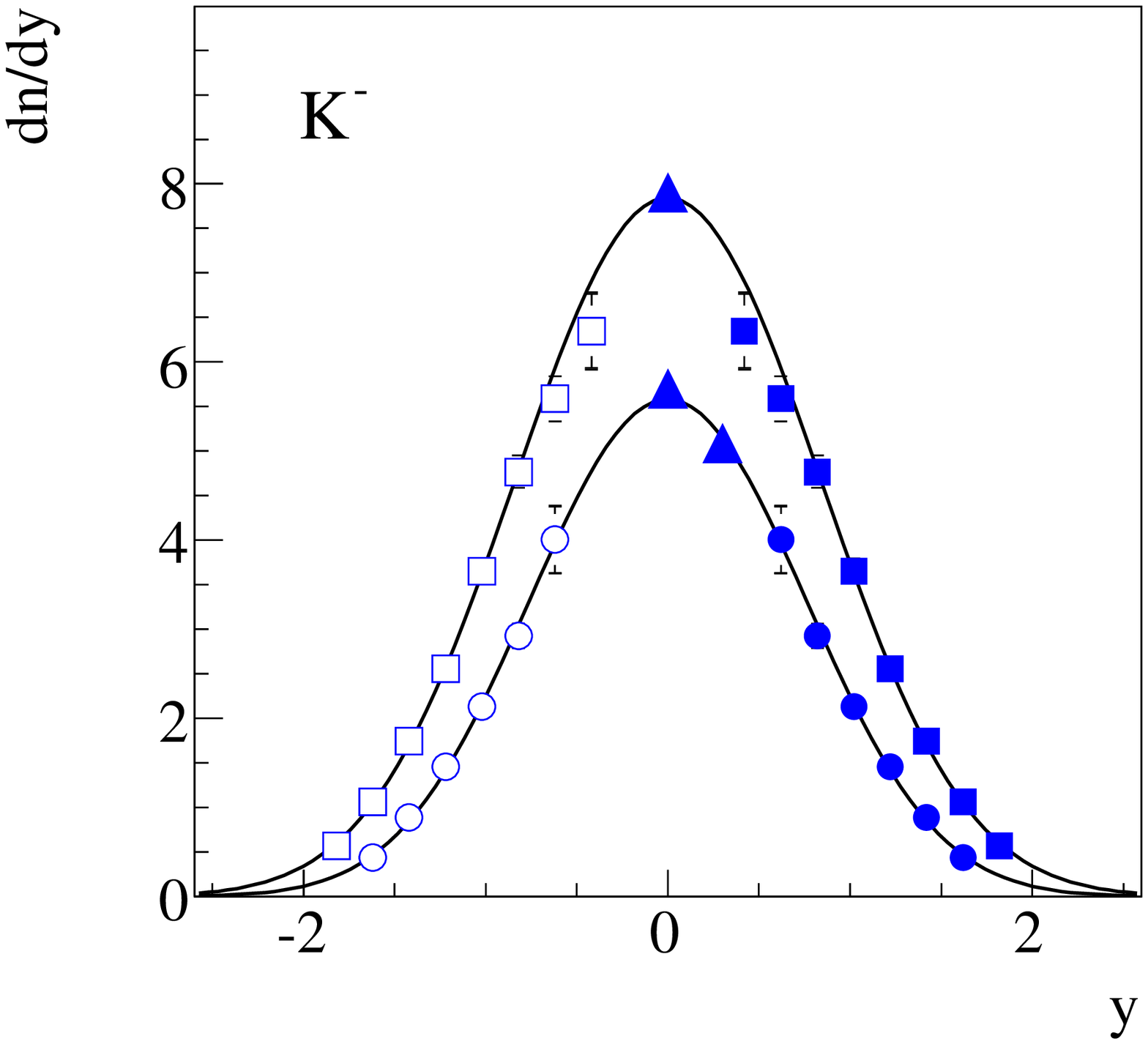}
       \includegraphics[width=0.33\linewidth]{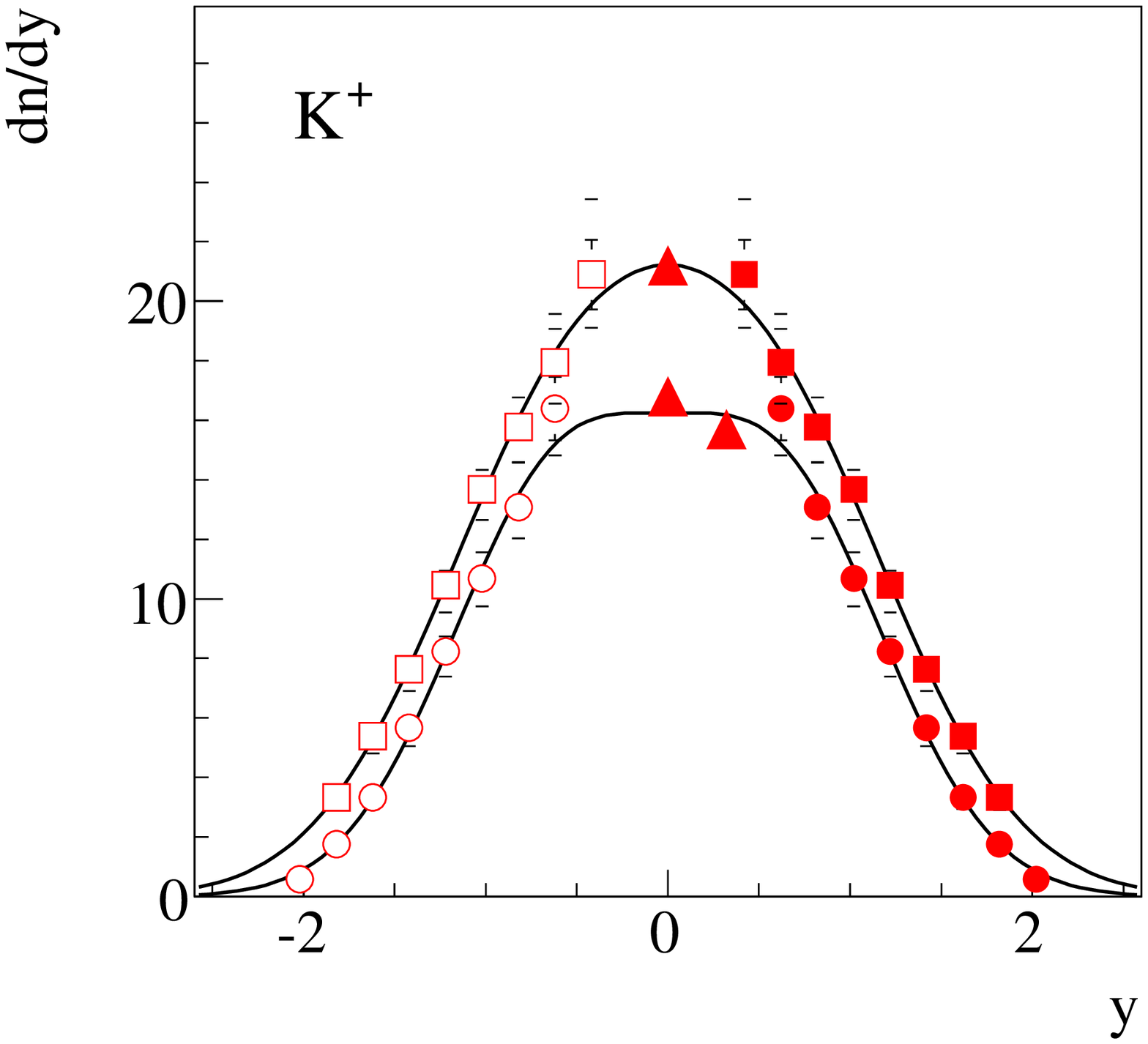} }
\caption{\label{fig2} 
  Rapidity distributions of $\pi^-$ (left), \kmin\ (middle) and
  \kplus\ mesons (right) produced in central Pb+Pb collisions at 20$A$
  (lower curves) and 30\agev (upper curves).  Squares and circles show
  the results of an analysis based on \dedx\ only whereas triangles
  are obtained from a \tofdedx\ analysis.  The closed symbols indicate
  measured points while the open points are reflected with respect to
  mid-rapidity. The lines indicate fits of \Eq{eq:twogauss} to the
  data.  The errors, which are often smaller than the symbol size, are
  statistical only.  The systematic errors on the measurements are
  about $\pm 5$\%.  }
\end{figure}
%
These distributions were obtained by summing the measured \mT\ spectra
and using the fitted exponential function \Eq{mt} to extrapolate to
full \mT.  For kaons the corresponding corrections are below 10\% for
most of the rapidity bins except for the first two and the last
rapidity bin from the \dedx\ only analysis in which not more than
about 50\% of the yield is in the measured region.  The corrections
for pions are negligible.  The rapidity spectra were parameterized by
the sum of two Gaussian distributions positioned symmetrically with
respect to mid-rapidity \cite{Afanasiev:2002mx}
\beq{eq:twogauss}
\frac {\der n} {\der y} = 
\frac { \langle n \rangle} { 2 \sqrt{2 \pi} \sigma }
\left\{
\exp\left[-\frac{1}{2}\left( \frac{y-y_0}{\sigma}\right)^2\right] +
\exp\left[-\frac{1}{2}\left( \frac{y+y_0}{\sigma}\right)^2\right]
\right\},
\eeq
where $\langle n \rangle$, $\sigma$ and $y_0$ are fit parameters.  The
results of the fits are indicated by the full lines in \Fi{fig2} and
the values of the parameters are given in \Ta{tab:results3}.  
Also given in this Table are the
mean
multiplicities which were obtained by integration of the fitted curves and
the mid-rapidity yields that were taken to be the maxima of the curves. 
\begin{table}[tbh]
\caption{\label{tab:results3} 
In the columns on the left are listed the inverse slope parameters $T$
obtained from a fit of \Eq{mt} to the \mT\ spectra at mid-rapidity,
together with the parameters $\sigma$ and $y_0$ from a fit of
\Eq{eq:twogauss} to the \pimin\ and \kplusmin\ rapidity spectra. Only
statistical errors are given.  The values of $\sigma$ and $y_0$ are
significantly correlated. In the columns on the right are listed the
full phase space yields $\ab{\piplusmin}$ and $\ab{\kplusmin}$
together with \dndy\ for \piplusmin\ and \kplusmin\ production at
mid-rapidity.  The first error is statistical, the second systematic.
Note that $\ab{\piplus}$ is not directly measured (see text).
}
\begin{ruledtabular}
\begin{tabular}{lll|lll}
 Parameter  & 20\agev\ & 30\agev & Yield &\multicolumn{1}{c}{20\agev}
            &\multicolumn{1}{c}{30\agev} \\
\hline

$T(\piplus)$\ \ (MeV)      & $167 \pm 2  $  & $175 \pm 2 $
  &  $\ab{\pimin}$         & $221 \pm 1 \pm 11$ 
                           & $274 \pm 1 \pm 14$             \\
$T(\pimin)$\ \  (MeV)      & $160 \pm 2  $  & $169 \pm 2 $
  &  $\ab{\piplus}$        & $190 \pm 1 \pm 9$ 
                           & $241 \pm 1 \pm 12$             \\
$T(\kplus)\,$\  (MeV)      & $219 \pm 5  $  & $232 \pm 5 $
  &  $\ab{\kplus}$         & $40.7 \pm 0.7 \pm 2.2$
                           & $52.9 \pm 0.9 ^{+3.0}_{-3.5}$      \\
$T(\kmin)\,$\   (MeV)      & $193 \pm 9  $  & $230 \pm 7 $
  &  $\ab{\kmin}$          & $10.3 \pm 0.1 \pm 0.2$ 
                           & $16.0 \pm 0.2 \pm 0.4$           \\
$\sigma(\pimin)$           & $0.837 \pm 0.007$ & $0.885 \pm 0.007\qquad$
  &  $\dndy(\pimin)$       & $84.8  \pm 0.4 \pm 4.2$ 
                           & $96.5  \pm 0.5 \pm 4.8$            \\
$\sigma(\kplus)$           & $0.601 \pm 0.012$ & $0.722 \pm 0.026$
  &  $\dndy(\piplus)$      & $72.9  \pm 0.3 \pm 3.6$ 
                           & $83.0  \pm 0.4 \pm 4.2$            \\
$\sigma(\kmin)$            & $0.642 \pm 0.035$ & $0.710 \pm 0.032$
  &  $\dndy(\kmin)$        & $5.58 \pm 0.07 \pm 0.11$
                           & $7.8  \pm 0.1 \pm 0.2$           \\
$y_0(\pimin)$              & $0.557 \pm 0.009$ & $0.624 \pm 0.009$
  &  $\dndy(\kplus)$       & $16.4 \pm 0.6  \pm 0.4$ 
                           & $21.2 \pm 0.8 ^{+1.5}_{-0.9}$      \\
$y_0(\kplus)$              & $0.606 \pm 0.014$ & $0.578 \pm 0.030$
  &                        &                   &                \\
$y_0(\kmin)$               & $0.34  \pm 0.06 $ & $0.37  \pm 0.05 $
  &                        &                   & 
\end{tabular}
\end{ruledtabular}
\end{table}
%
The mean multiplicity of $\pi^+$ mesons and their mid-rapidity yields
were calculated by scaling $\langle \pi^- \rangle$ with the
$\pi^+/\pi^-$ ratio measured at mid-rapidity in the \tofdedx\ 
analysis.  These ratios were found to be $(0.86,0.88,0.90,0.91,0.93)$
for the data measured at $(20,30,40,80,158)$\agev. The ratios at the
latter three energies were used to recalculate the published values of
the $\langle \pi^+ \rangle$ and the $\pi^+$ mid-rapidity yields at
these energies.  The recalculated $\pi^+$ multiplicities differ by
several percent from the ones published in~\cite{Afanasiev:2002mx}.

\section{Review of the energy dependence}

In this section, the new results on $\pi$ and K production at 20$A$ and
30\agev\ will be discussed together with published measurements at
lower (AGS) and higher (SPS, RHIC) energies and compared to the
corresponding data from $\proton +\proton (\pbar)$ interactions.
Model calculations, which are shown by the curves in the figures
below, will be discussed in the next section.

In \Fi{edep_pions}
%
\begin{figure}[htb]
\includegraphics[width=0.49\linewidth]{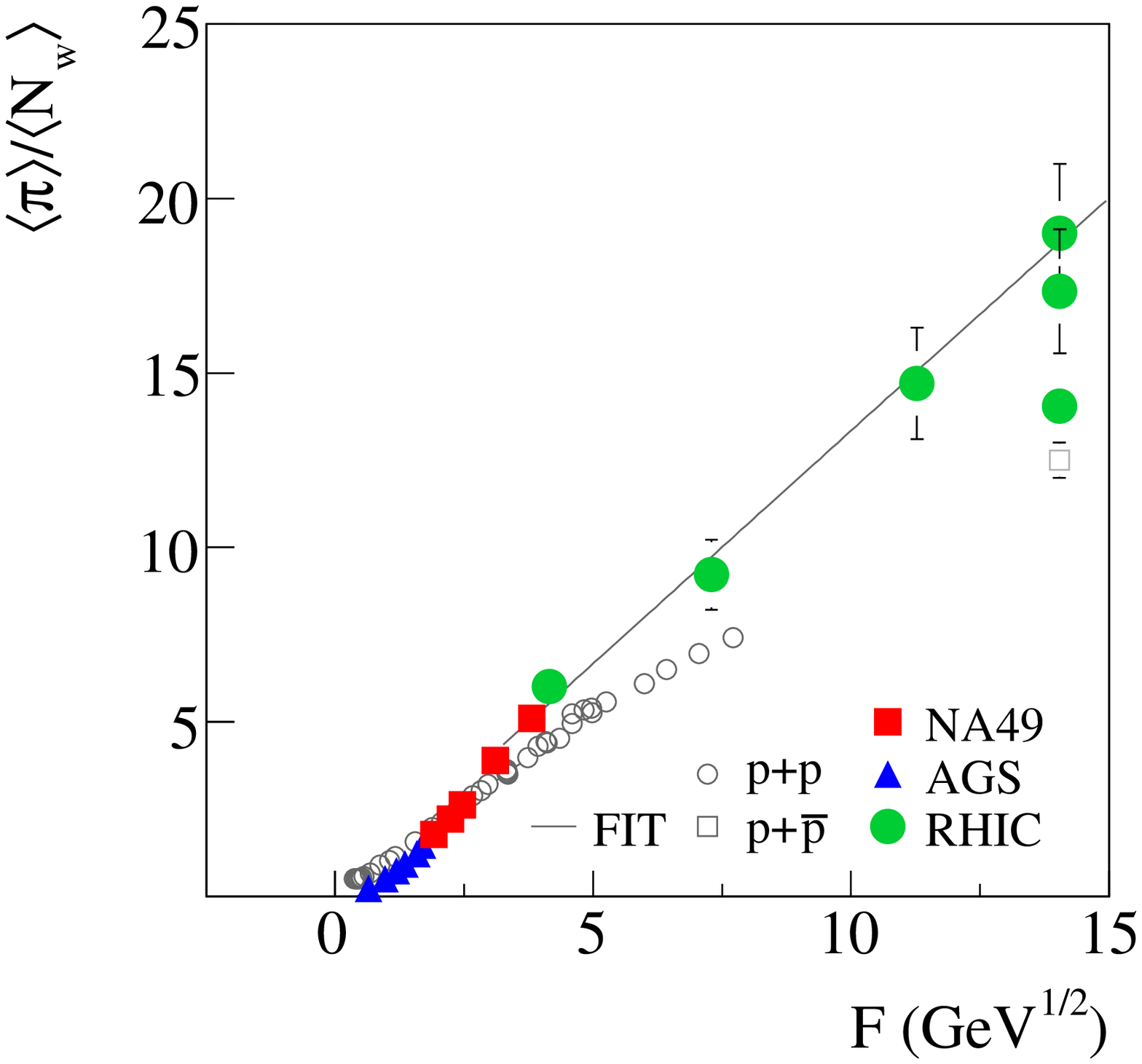}
\includegraphics[width=0.49\linewidth]{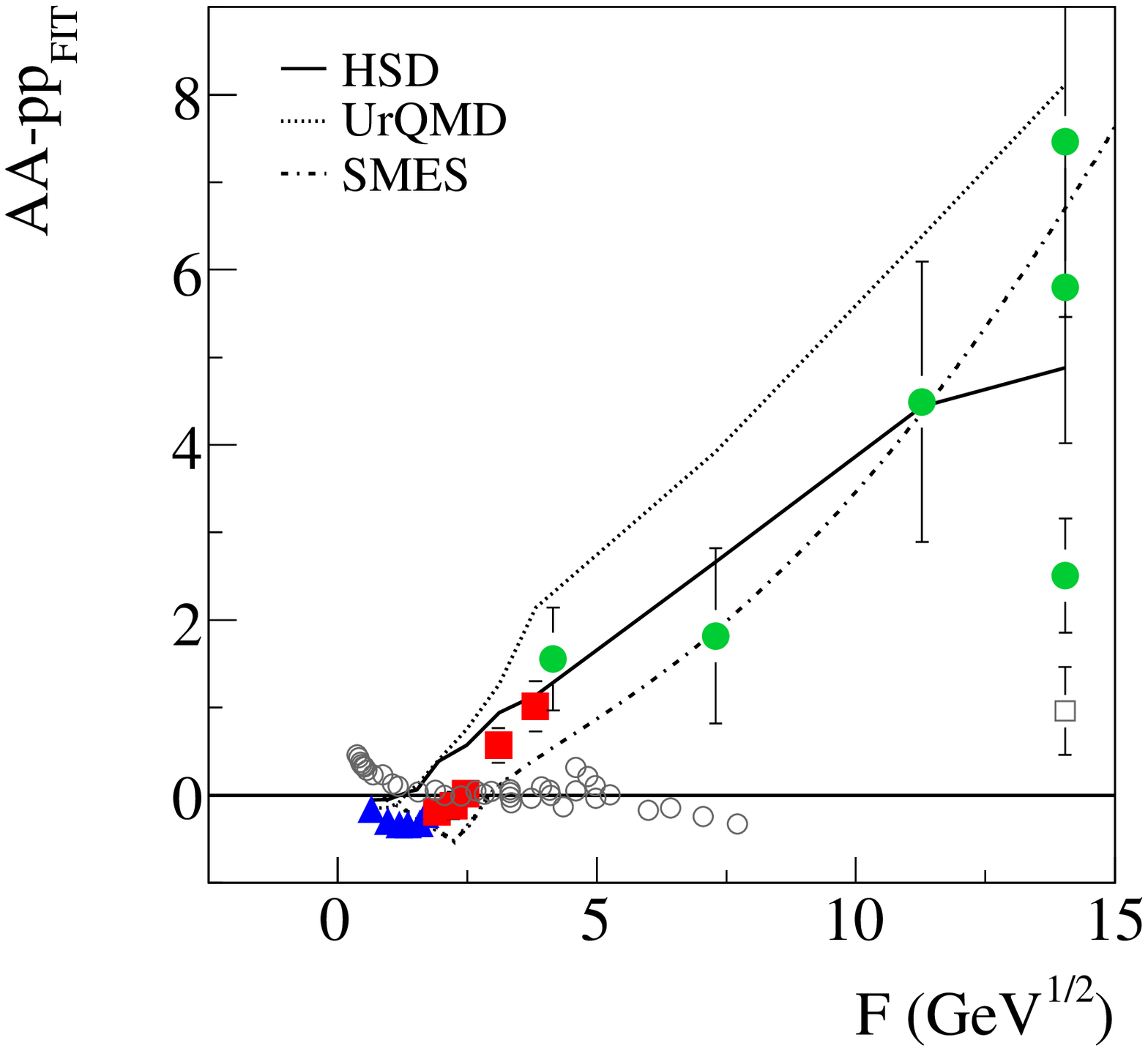}
\caption{\label{edep_pions}
  Left: Energy dependence of the mean pion multiplicity per wounded
  nucleon measured in central Pb+Pb and Au+Au~\cite{AGS,RHIC}
  collisions (full symbols), compared to the corresponding results
  from $\proton +\proton (\pbar)$ reactions (open circles).  Right:
  Energy dependence of the difference between the measured mean pion
  multiplicity per wounded nucleon and a parametrization (see text) of
  the $\proton + \proton$ data.  The meaning of the full and open
  symbols is the same as in the left-hand plot. The lines show various
  model predictions described in the text.  }
\end{figure}
%
is shown the mean pion multiplicity $\ab{\pi} = 1.5\,
(\ab{\pi^+} + \ab{\pi^-})$ per wounded nucleon
\nwound\ as a function of the collision energy, expressed by Fermi's
measure \cite{Fe:50}
\[
   F \equiv \left[ \frac{(\sqrt{s_{NN}} - 2 m_N)^3}
   {\sqrt{s_{NN}}} \right]^{1/4} ,
\]
where $\sqrt{s_{NN}}$ is the centre-of-mass energy per nucleon-nucleon
pair and $m_N$ the nucleon rest mass.  The use of $F$ in
Fig.~\ref{edep_pions} is motivated by model expectations, as will be
discussed below.

In \Fi{edep_pions} the measurements by NA49 are compared to results
from the AGS~\cite{AGS} and RHIC~\cite{RHIC} on central nucleus-nucleus
collisions.  The lowest point at the top RHIC energy is obtained from
the BRAHMS measurements of identified hadron spectra.  The middle and
highest points are results from, respectively, BRAHMS and PHOBOS on
the charged hadron mean multiplicity corrected for non-pion
contributions by use of the BRAHMS identified hadron yields.  The RHIC
points at the lower energies are from PHOBOS, again corrected for the
non-pion contribution.  The results from $\proton +\proton (\pbar)$
interactions are shown by the open symbols. For ease of comparison
these data were fitted to the parameterization
\beq{eq:pivsf}
  \frac{\ab{\pi}}{\nwound} = a + b\, F + c\, F^2,
\eeq
resulting in $a = -0.44 \pm 0.06$, $b = 1.32 \pm 0.03$~GeV$^{-1/2}$
and $c = -0.033 \pm 0.004$~GeV$^{-1}$.  Up to the top SPS energy the
mean pion multiplicity in $\proton + \proton$ interactions is
approximately proportional to $F$.  A fit 
of \Eq{eq:pivsf}
in the range $2 < F <
5$~GeV$^{1/2}$ with $a = c = 0$ yielded a value of $b
= 1.063 \pm 0.003$ GeV$^{-1/2}$.

For central Pb+Pb and Au+Au collisions the energy dependence is more
complicated as is seen in the right panel of \Fi{edep_pions} where the
difference between the data and the $\proton + \proton$
parameterization Eq.~\eq{eq:pivsf} is plotted.  Below 40\agev\ the ratio
$\ab{\pi}/\nwound$ is lower in A+A collisions than in p+p
interactions (pion suppression) while at higher energies this ratio is
larger in A+A collisions than in $\proton + \proton$ interactions
(pion enhancement).  The transition from pion suppression to pion
enhancement is clearly demonstrated in the figure.  A linear fit for
$F < 1.85$~GeV$^{1/2}$ using Eq.~\eq{eq:pivsf} with $c = 0$ gave $a = -0.45
\pm 0.05$ and $b = 1.03 \pm 0.05$~GeV$^{-1/2}$ .  The slope parameter
fitted in the range $F > 3.5$~GeV$^{1/2}$ was $b = 1.33 \pm 0.03$
where the lowest point at the top RHIC energy was excluded from the
fit.  Thus, in the region 15--40\agev\ between the highest AGS and the
lowest SPS energy the slope increases by a factor of about 1.3.
 
\begin{figure}[htb]
\includegraphics[width=0.49\linewidth]{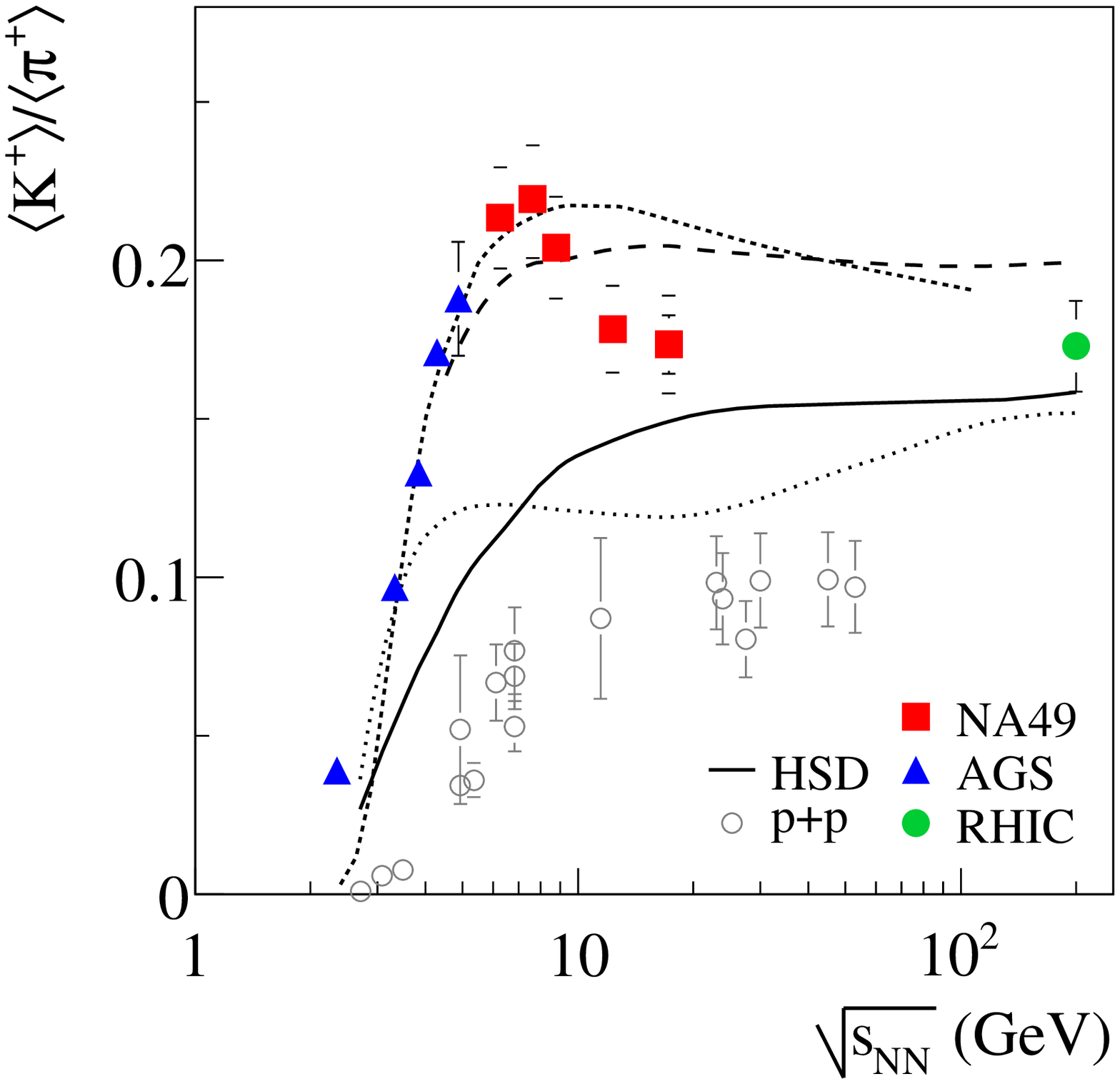}
\includegraphics[width=0.49\linewidth]{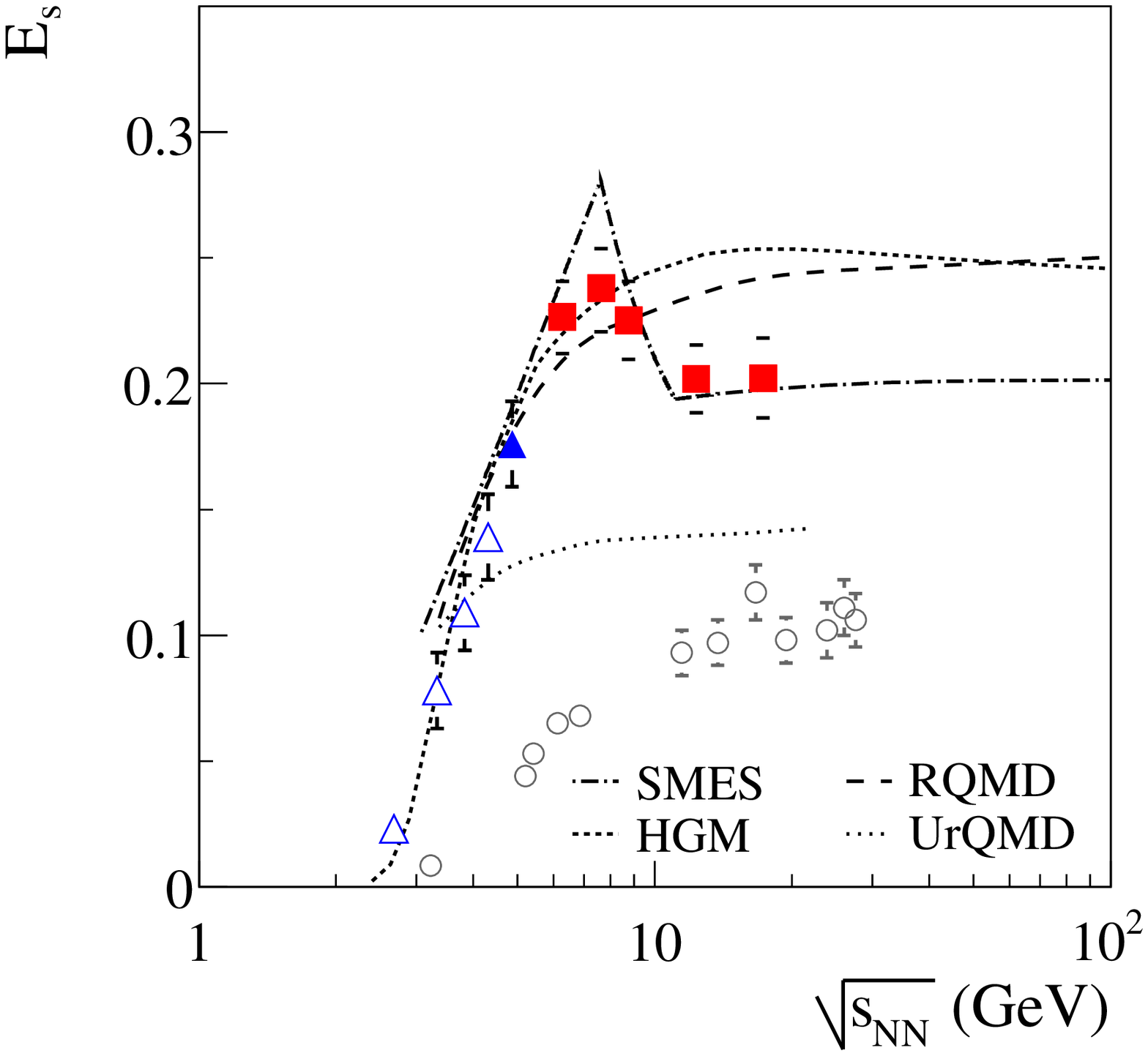}
\caption{\label{edep_strangeness}
  Left: Energy dependence of the $\ab{\kplus}/\ab{\pi^+}$ ratio
    measured in central Pb+Pb and Au+Au~\cite{AGS,RHIC}) collisions
    (full symbols) compared to the corresponding results from $\proton
    +\proton (\pbar)$ reactions (open cirles). Right: Energy
    dependence of the relative strangeness production as measured by
    the \eejes\ ratio (see text) in central Pb+Pb and Au+Au collisions
    (full symbols) compared to results from $\proton +\proton (\pbar)$
    reactions (open circles).  The curves in the figures show
    predictions of various models described in the text.  }
\end{figure}

Fig.~\ref{edep_strangeness}
%
%
shows the full phase space ratios $\ab{\kplus}/\ab{\pi^+}$ and $\eejes
= (\ab{\Lambda} + \ab{{\rm K} +\kbar})/\ab{\pi}$ as a function of
collision energy in the left and right panels, respectively.  Kaons
are the lightest strange hadrons and $\ab{\kplus}$ accounts for about
half of all the anti-strange quarks produced in Pb+Pb collisions at
AGS and SPS energies (a detailed explanation is given below).  In the
\eejes\ ratio all main carriers of strange and anti-strange quarks are
included. 
The values for Pb+Pb and Au+Au collisions were calculated using
data from this paper and 
Refs.~\cite{Afanasiev:2002mx,Anticic:2003ux,AGS,Mitrovski:2006js}.
The neglected contribution of $\overline{\Lambda}$ and other
hyperons and anti-hyperons is about 10\% at SPS energies.  Both the
$\ab{\kplus}/\ab{\pi^+}$ and \eejes\ ratios are approximately, 
within 5\% at SPS energies, proportional to
the ratio of total multiplicity of $s$ and $\overline{s}$ quarks to
the multiplicity of pions.  It should be noted that the
$\ab{\kplus}/\ab{\pi^+}$ ratio is expected to be similar (within about
10\%) for $\proton + \proton$, $\neutron + \proton$ and $\neutron +
\neutron$ interactions at 158\agev~\cite{hansen}, whereas the \eejes\ 
ratio is independent of the isospin of nucleon-nucleon interactions.
Calculating the \eejes\ ratio for the $\proton + \proton$ data 
\cite{GaRo} is
significantly more precise than taking the ratio
$\ab{\kplus}/\ab{\pi^+}$.

It is seen from \Fi{edep_strangeness} that a steep increase of both
ratios in the  AGS energy region is followed by a turnover and a
decrease around 30\agev.  The BRAHMS measurements at the top RHIC
energy~\cite{RHIC} indicate that the $\ab{\kplus} / \ab{\pi^+}$ ratio
stays nearly constant starting from the top SPS energy.  The RHIC
results for the \eejes\ ratio are not available because the total
multiplicity of $\Lambda$ hyperons is not measured.
For comparison the results from $\proton + \proton$
interactions~\cite{GaRo} are also plotted in \Fi{edep_strangeness}.
These data show a monotonic increase with increasing energy.

\begin{figure}[htb]
\includegraphics[width=0.49\linewidth]{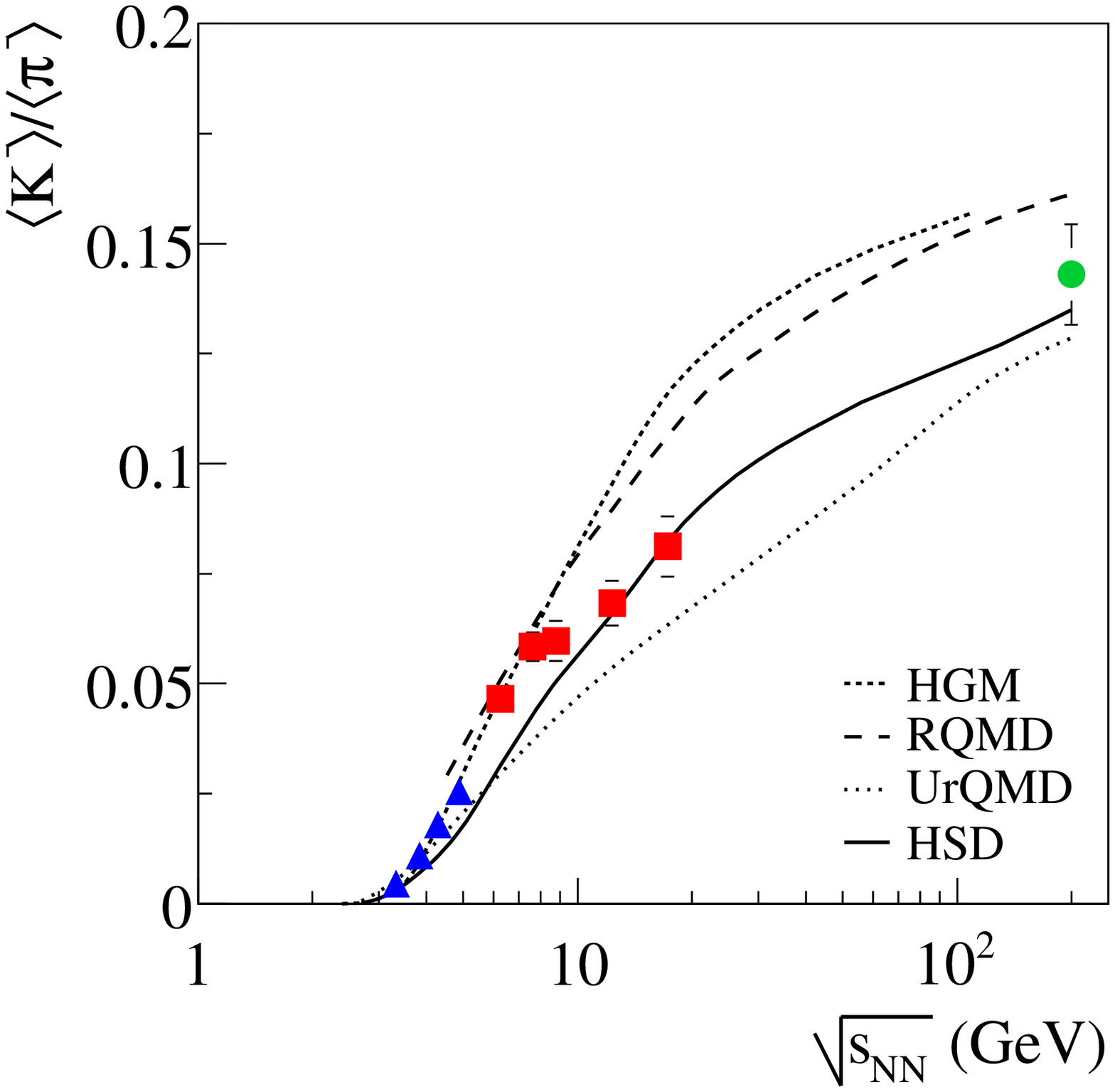}
\includegraphics[width=0.46\linewidth]{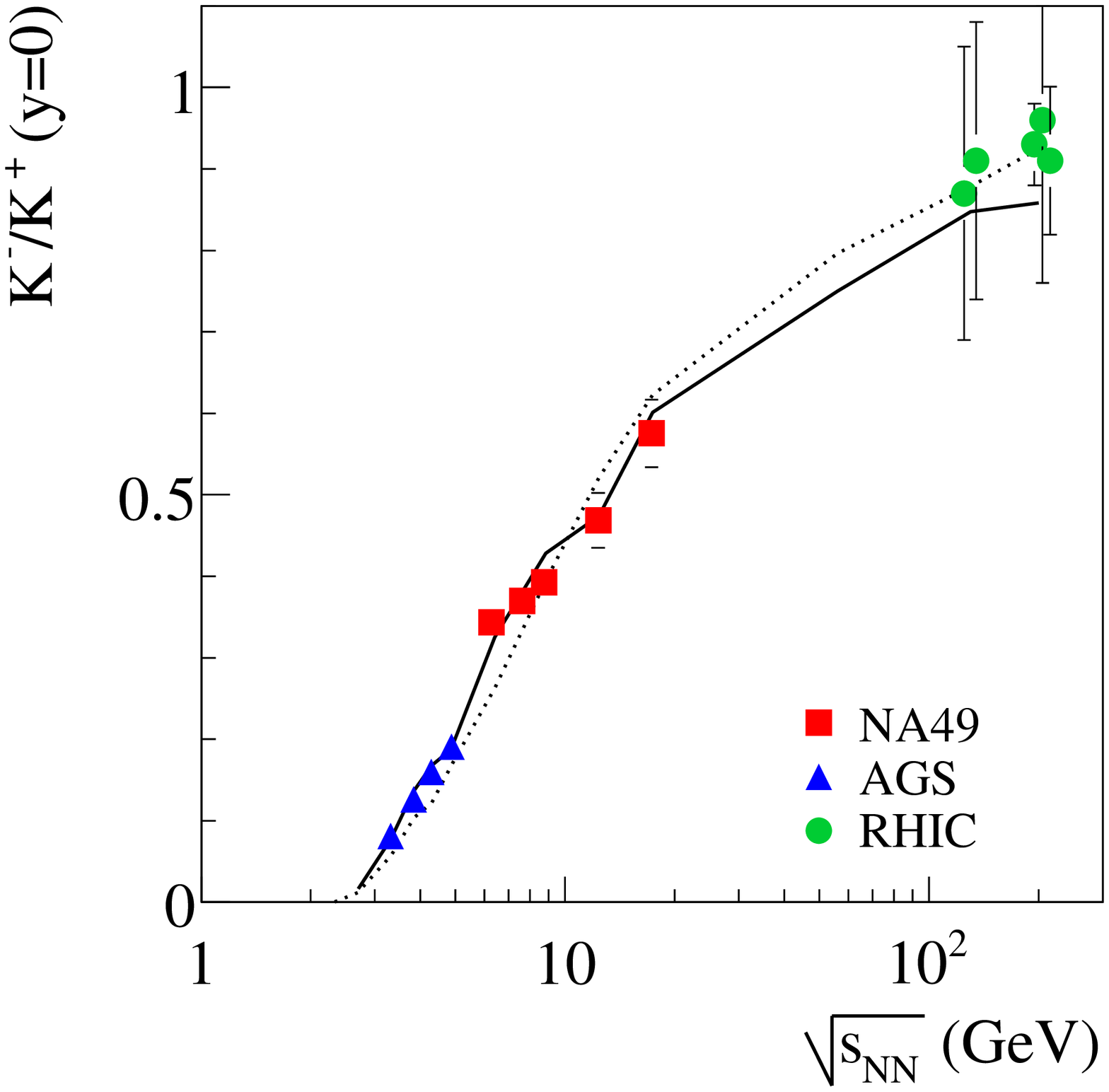}
\caption{\label{edep_kmkp}
  Left: Energy dependence of the ratio $\ab{\kmin} / \ab{\pi^-}$
  measured in central Pb+Pb and Au+Au~\cite{AGS,RHIC} collisions (full
  symbols).  Right: Energy dependence of the ratio of mid-rapidity
  yields of \kmin\ and \kplus\ mesons. The curves in the figures show
  model predictions described in the text.
  }
\end{figure}

The ratios $\ab{\kmin} / \ab{\pi^-}$ and
$\kmin / \kplus $ at mid-rapidity (see below for details) 
increase monotonically with increasing
collision energy, as can be seen from Fig.~\ref{edep_kmkp}.
%
%
The difference between the dependence of the \kplus\ and \kmin\ yields
on collision energy can be attributed to their different sensitivity
to the baryon density.  K$^+$ and K$^0$ carry a dominant fraction of
all produced $\overline{s}$-quarks exceeding 95\% in Pb+Pb collisions
at 158\agev\ if open strangeness is considered. Because $\ab{\kplus}
\cong \ab{{\rm K}^0}$ in approximately isospin symmetric collisions of
heavy nuclei, the \kplus\ yield is nearly proportional to the total
strangeness production and only weakly sensitive to the baryon
density.  As a significant fraction of $s$-quarks (about 50\% in
central Pb+Pb collisions at 158\agev) is carried by hyperons, the
number of produced anti-kaons \kmin and $\kbar^0$ is sensitive to both
the strangeness yield and the baryon density.
 
In \Fi{edep_kaons_mid}
%
\begin{figure}[htb]
\includegraphics[width=0.49\linewidth]{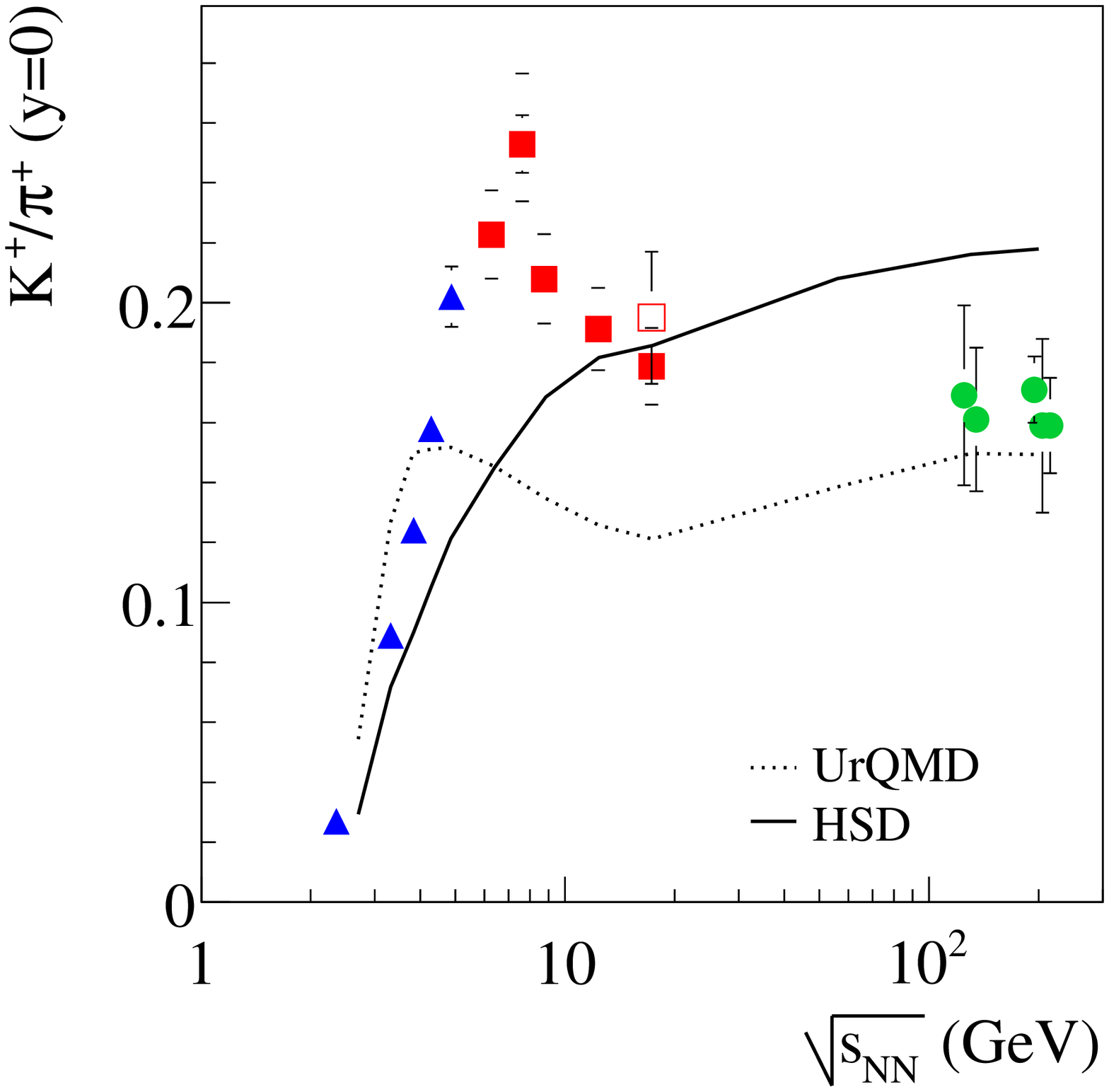}
\includegraphics[width=0.49\linewidth]{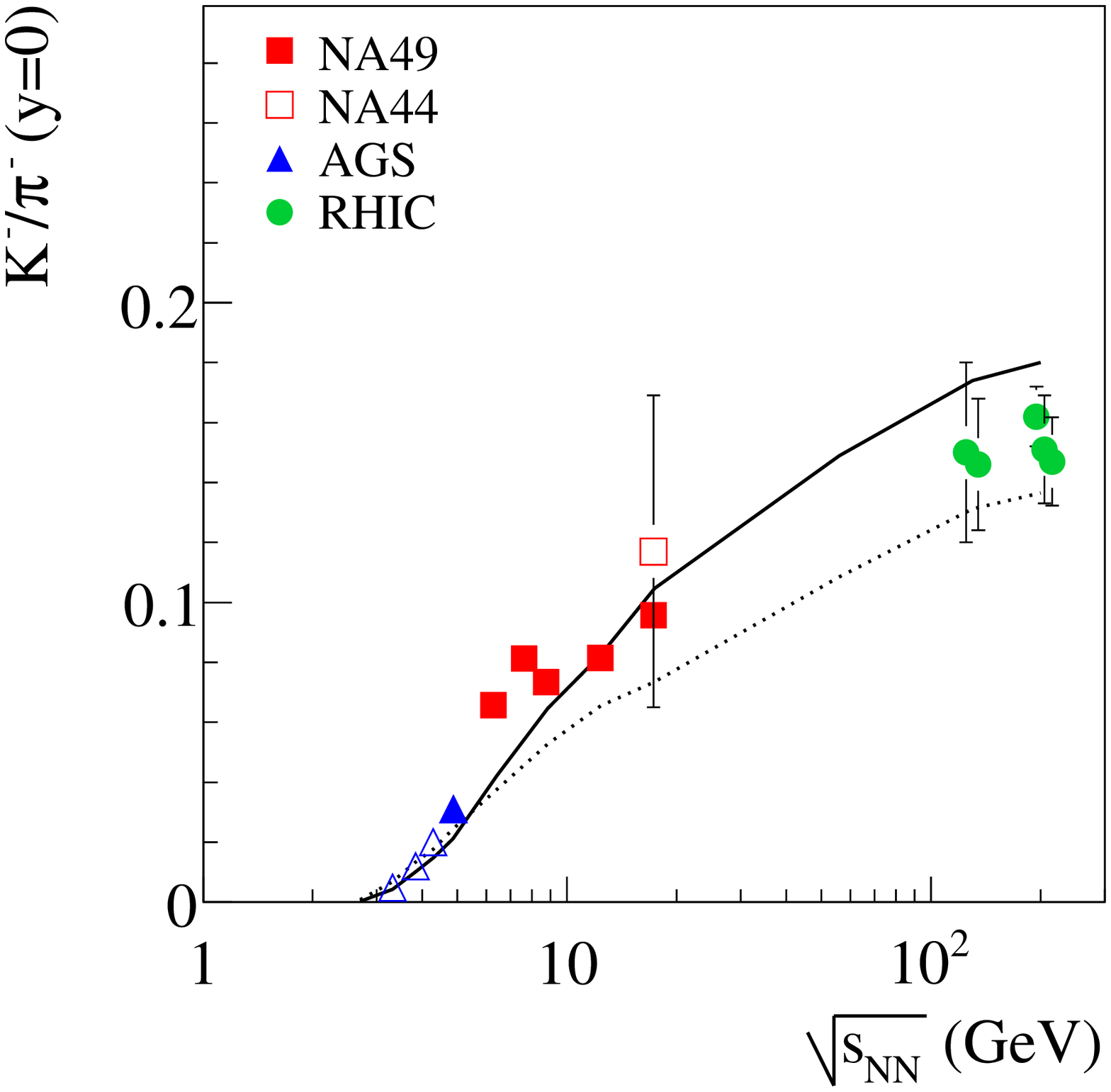}
\caption{\label{edep_kaons_mid} 
  Energy dependence of the ratio $\kplus / \pi^+ $ (left) and $\kmin /
  \pi^- $ (right) at mid-rapidity measured in central Pb+Pb and
  Au+Au~\cite{AGS,RHIC} collisions.  The NA44 data are taken
  from~\cite{NA44}.
  }
\end{figure}
%
the mid-rapidity ratios $\kplus / \pi^+$ and $\kmin / \pi^-$ for
central Pb+Pb and Au+Au collisions are shown as a function of the
collision energy.  All NA49 mid-rapidity results presented here are
determined from the fits of the rapidity spectra to \Eq{eq:twogauss}.
The statistical and systematic errors were calculated taking into
account all correlations between the fitted parameters. It is seen
from \Fi{edep_kaons_mid} that the NA44 data point~\cite{NA44} is
consistent with the NA49 results and that the STAR and PHENIX
measurements at the top RHIC energies are in agreement with the trend
seen in the SPS results. Comparison with the left-hand plots of
\Fi{edep_strangeness} and~\ref{edep_kmkp} shows that the energy
dependence of the mid-rapidity kaon to pion ratios is similar to that of
the corresponding ratios measured in full phase-space.

\begin{figure}[htb]
\includegraphics[width=0.49\linewidth]{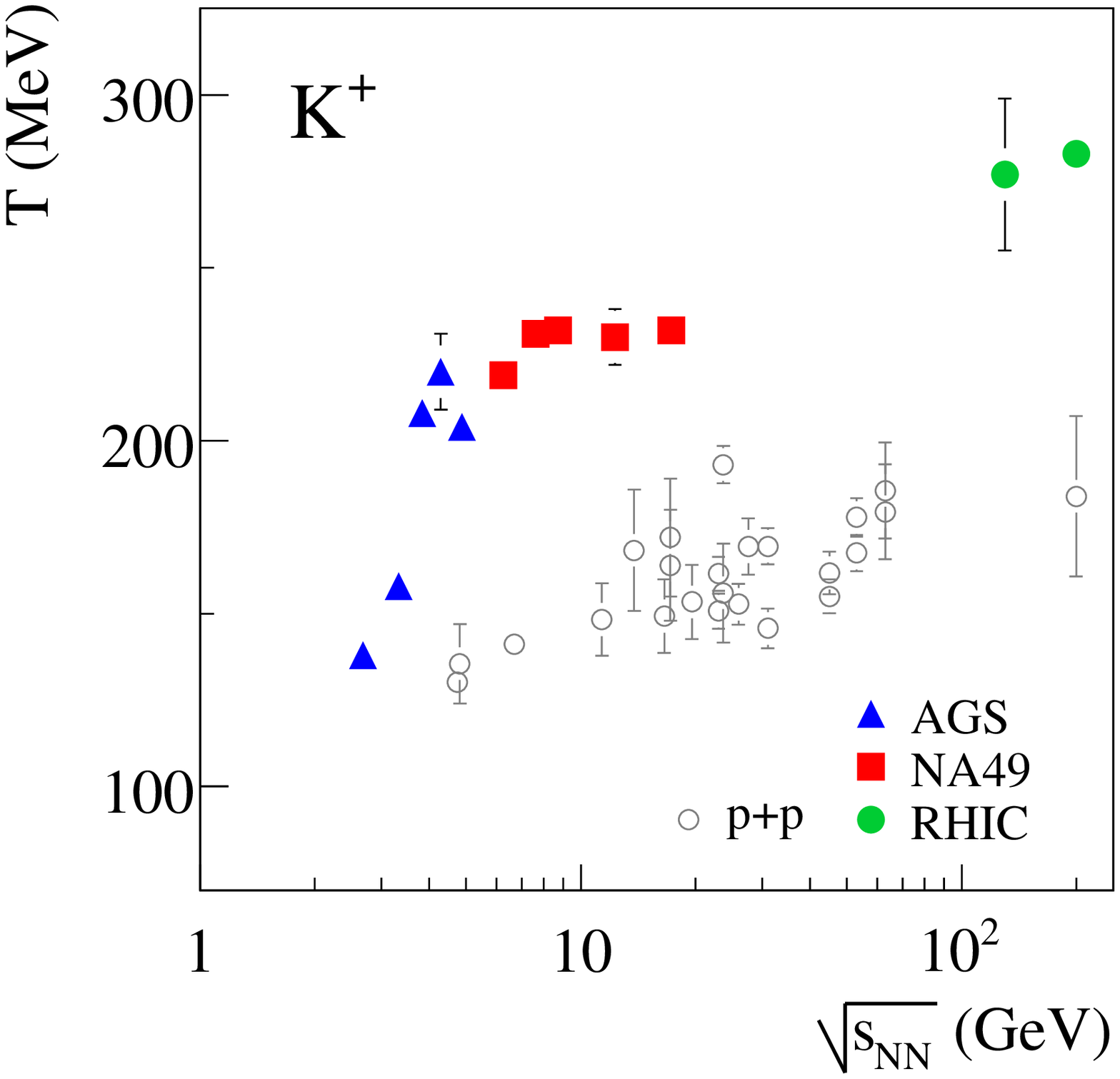}
\includegraphics[width=0.49\linewidth]{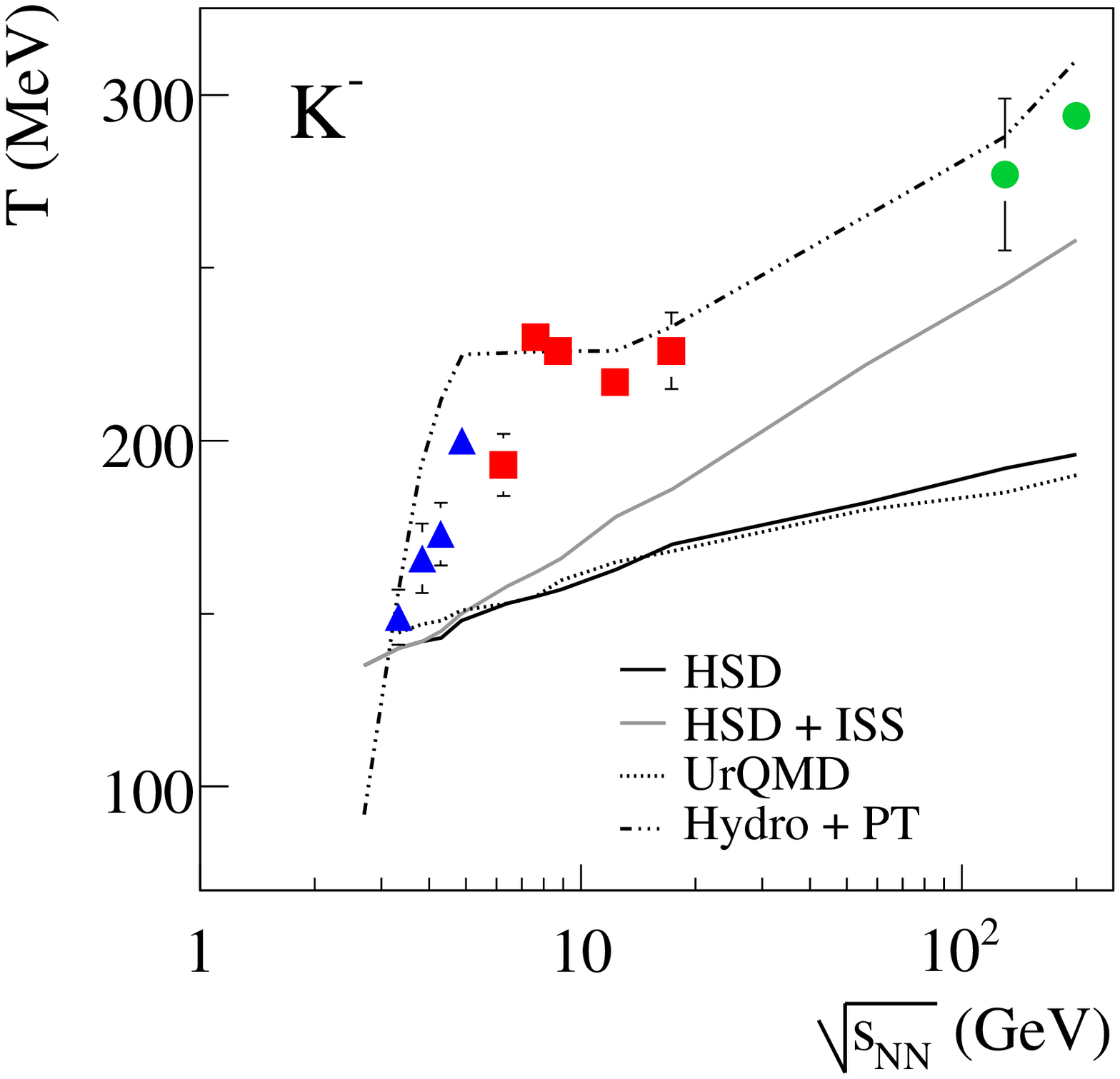}
\caption{\label{edep_slopes}
  Energy dependence of the inverse slope parameter $T$ of the
  transverse mass spectra of \kplus\ (left) and \kmin\ mesons (right)
  measured at mid-rapidity in central Pb+Pb and Au+Au~\cite{AGS,RHIC}
  collisions. The \kplus\ slope parameters are compared to those from
  $\proton +\proton(\pbar)$ reactions~\cite{kliemant} in the left-hand
  plot (open circles). The curves in the right-hand plot represent
  predictions from various models described in the text.
  }
\end{figure}

Fig.~\ref{edep_slopes}
%
%
presents the energy dependence of the inverse slope parameter $T$ of
the transverse mass spectra of \kplus\ (left panel) and \kmin\ mesons
(right panel) produced in central Pb+Pb and Au+Au collisions.  One
observes a plateau at SPS energies which is preceded by a steep rise
of $T$ measured at the AGS~\cite{AGS} and followed by an indication of
a further increase of the RHIC data~\cite{RHIC}.  Although the scatter
of data points is large, $T$ appears to increase smoothly in $\proton
+\proton(\pbar)$ interactions~\cite{kliemant} as shown in the left
panel of \Fi{edep_slopes}.

The transverse mass spectra of pions and protons are non-exponential
such that the inverse slope parameter depends on the transverse mass
interval used in the fit.  The mean transverse mass $\ab{\mT} - m$
provides an alternative characterization of the \mT\ spectra that
avoids this problem. It was calculated from multi-parameter fits to
the measured data in the interval $\mT - m \leq 2.0$~GeV.
For pions either a sum of two exponentials or a power law function was
used, which both yield a good description of the data.  In case of
kaons either a single exponential or a power law was used.  The final
$\ab{\mT} - m$ values are the average of the result from both fit
methods.  The error is the quadratic sum of the statistical error
and a systematic contribution that is
estimated from the differences in the results obtained with the two
different fit functions.  The values of $\ab{\mT}$ at AGS and RHIC
energies were calculated from the spectra published
in~\cite{AGS,RHIC}.  The measurements of $\ab{\mT}$ for p and \pbar\ 
at the SPS were taken from~\cite{NA49_ppbar}.  The energy dependence
of $\ab{\mT}$ for pions, kaons and protons is shown in
\Fi{edep_mt}.
%
\begin{figure}[htb]
\includegraphics[width=0.90\linewidth]{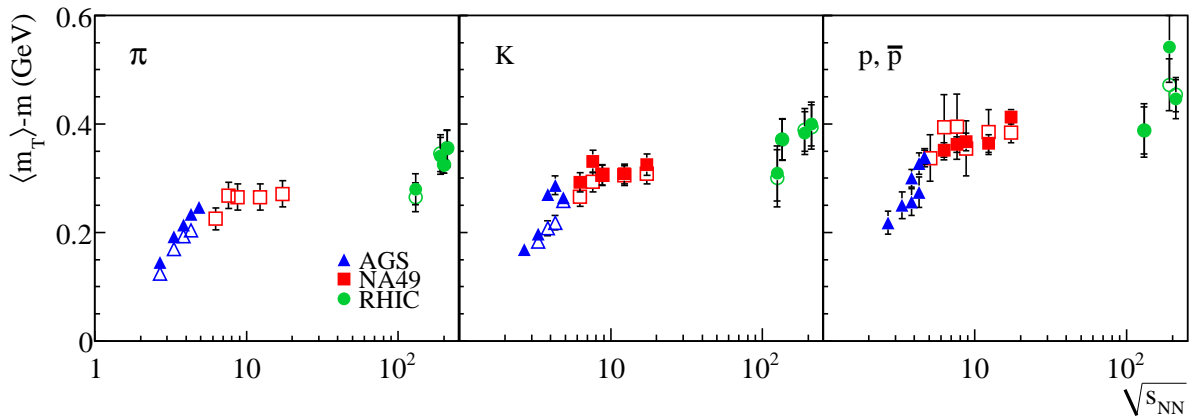}
\caption{\label{edep_mt}
  Energy dependence of the mean transverse mass $\ab{\mT}$ measured at
  mid-rapidity in central Pb+Pb an Au+Au collisions for $\pi^{\pm}$
  (left) and \kplusmin\ (middle).  For completeness, the previously
  published results on p and \pbar~\cite{NA49_ppbar} are shown in the
  right panel. In the plots, positively (negatively) charged hadrons
  are indicated by the full (open) symbols.  The results from AGS and
  RHIC are taken from~\cite{AGS,RHIC}.  }
\end{figure}
%
The results show that the approximate energy independence of
$\ab{\mT}$ in the SPS energy range is a common feature for all
particles investigated.

In conclusion, rapid changes in the energy dependence of pion and kaon
production properties are observed which all seem to coincide in the
low SPS energy range of 20--30\agev.  This suggests that a common
underlying physics process is responsible for these changes.

\section{Discussion of model explanations}

In this section the energy dependence of pion and strangeness
production properties will be discussed within various approaches to
nucleus-nucleus collisions and compared with published model
predictions.

It was suggested in~\cite{GaRo,Ga:95} that a transition to a
deconfined state of matter may cause anomalies in the energy
dependence of pion and strangeness production in nucleus-nucleus
collisions.  This led to the formulation of the Statistical Model of
the Early Stage (SMES)~\cite{Ga:95,GaGo} which is based on the
assumption that the system created at the early stage (be it confined
matter or a QGP) is in equilibrium and that a transition from a
reaction with purely confined matter to a reaction with a QGP at the
early stage occurs when the transition temperature \Tc\ is reached.
For \Tc\ values of 170--200~MeV the transition region extends from 15$A$
to 60\agev~\cite{GaGo}.  Assuming the generalized Fermi-Landau
conditions~\cite{GaGo,Fe:50,FeLa} for the early stage of
nucleus-nucleus collisions and a proportionality of the pion
multiplicity to the early stage entropy, the ratio $\ab{\pi} /
\nwound$ increases linearly with the Fermi measure $F$ outside the
transition region.  The slope parameter is proportional to
$g^{1/4}$~\cite{Ga:95}, where $g$ is the effective number of internal
degrees of freedom at the early stage.  In the transition region a
steepening of the pion energy dependence is due to the activation of a
large number of partonic degrees of freedom.  This is, in fact,
observed in the data on central Pb+Pb and Au+Au collisions, where the
steepening starts at about 20\agev, as shown in \Fi{edep_pions}.  The
linear dependence of $\ab{\pi} / \nwound$ on $F$ is approximately
obeyed by the data at lower and higher energies (including RHIC).  An
increase of the slope by a factor of about 1.3 is measured, which
corresponds to an increase of the effective number of internal degrees
of freedom by a factor of 1.3$^4$ $\cong$ 3, within the
SMES~\cite{Ga:95}.

The $\ab{\kplus} / \ab{\pi^+}$ and \eejes\ ratios are roughly
proportional to the total strangeness to entropy ratio which in the
SMES model is assumed to be preserved from the early stage till
freeze-out.  At low collision energies the strangeness to entropy
ratio increases steeply with collision energy, due to the low
temperature at the early stage ($T < \Tc$) and the high mass of the
strangeness carriers in the confined state (the kaon mass,
for instance, is 500~MeV).  When the transition to a QGP is crossed
($T > \Tc$), the mass of the strangeness carriers is significantly
reduced to the strange quark mass of about 100~MeV.  Due to the low
mass $m < T$, the strangeness yield becomes (approximately)
proportional to the entropy, and the strangeness to entropy (or pion)
ratio is independent of energy.  This leads to a decrease in the
energy dependence from the larger value for confined matter at \Tc\ to
the QGP value. Thus the measured non-monotonic energy dependence of
the strangeness to entropy ratio is followed by a saturation at the
QGP value. Such anomalous energy dependence can indeed be seen
in~\Fi{edep_strangeness} and is, within the SMES, a direct consequence
of the onset of deconfinement taking place at about 30\agev.

In the mixed phase region the early stage pressure and temperature are
independent of the energy density~\cite{VanHove:1982vk}.
Consequently, within the SMES model this should lead to the weakening
of the increase with energy of the inverse slope parameter $T$ or,
equivalently, the mean transverse mass $\ab{\mT}$ in the SPS energy
range~\cite{Gorenstein:2003cu}.  This qualitative prediction is
confirmed by the results shown in Figs.~\ref{edep_slopes} and
\ref{edep_mt}. Moreover, recent hydrodynamic
calculations~\cite{brasil} that model both the deconfined and hadronic
phases provide a quantitative description of the data as shown by the
dashed-dotted curve in \Fi{edep_slopes}.

Several other analyses of the energy dependence of hadron production
properties in central Pb+Pb and Au+Au collisions within various
theoretical approaches support the hypothesis that the onset of
deconfinement is located at the low SPS energies.  In particular such
a result was obtained from studies of hadron yields within a
non-equilibrium hadron gas model~\cite{rafelski} and using the
momentum integrated Boltzmann equation for a description of the time
evolution of the relative strangeness yield~\cite{nayak}.  Furthermore
it was deduced from the experimentally measured rapidity spectra that
within Landau's hydrodynamical model the sound velocity at the early
stage of the reaction has a minimum at about 30\agev~\cite{bleicher}.
A minimum of the sound velocity is expected to occur in the phase
transition domain.  Moreover, a simultaneous analysis of the two-pion
correlation function and the transverse mass spectra found a plateau
in the averaged phase-space density at SPS energies which may be
associated with the onset of deconfinement~\cite{sinyukov}.

Numerous models have been developed to explain hadron production in
reactions of heavy nuclei without explicitly invoking a transient QGP
phase.  The simplest one is the statistical hadron gas
model~\cite{Ha:94} where independent of the collision energy the
hadrochemical freeze-out creates a hadron gas in equilibrium.  The
temperature, the baryon chemical potential and the hadronization
volume are free parameters of the model and are fitted to the data at
each energy.  In this formulation, the hadron gas model cannot 
predict the energy dependence of hadron production so that an
extension of the model was proposed, in which the values of the
temperature and baryon chemical potential evolve smoothly with
collision energy~\cite{Cl:01}.  The energy dependence of the \eejes\ 
ratio calculated within this extended hadron gas model is compared to
the experimental results in \Fi{edep_strangeness}.  By construction,
the prevailing trend in the data is reproduced by the model but the
decrease of the ratio between 30$A$ and 80\agev\ is not well described.
The measured strangeness to pion yield in central Pb+Pb collisions at
158\agev\ is about 25\% lower than the expectation for the fully
equilibrated hadron gas~\cite{Cl:01,Be:98}.  Obviously the
non-equilibrium hadron gas models~\cite{Be:03,Be:05,rafelski} with the
parameters fitted separately to the data at each energy describe the
experimental results significantly better.  Interestingly, it was
found in~\cite{Cl:05} that the transition from baryon to meson
dominated freeze-out conditions happens to be located at low SPS
energies.

Dynamical models of A+A collisions, such as RQMD~\cite{RQMD},
UrQMD~\cite{URQMD} and HSD~\cite{HSD} treat the initial
nucleon-nucleon interactions within a string-hadronic framework.  In
addition these models include effects such as string-string
interactions and hadronic re-scattering which are expected to be
relevant in A+A collisions.  The predictions of the
RQMD~\cite{RQMD,RQMD1}, UrQMD~ \cite{URQMD,URQMD1,Bratkovskaya:2004kv}
and HSD~\cite{Bratkovskaya:2004kv} models are shown in
Figs.~\ref{edep_pions}, \ref{edep_strangeness}, \ref{edep_kmkp},
\ref{edep_kaons_mid} and \ref{edep_slopes}.  It is seen that all these
models, like the hadron gas model, fail to describe the rapid change
of the hadron production properties with collision energy in the low
SPS energy range.  It was recently shown that the maximum in relative
strangeness production can be reproduced by invoking an unusually long
lifetime of the fireball at low SPS energies which decreases with the
collision energy~\cite{tomasik}. This assumption is however difficult
to justify with the dynamical models of the collision
process~\cite{URQMD,HSD} and the results on the energy dependence of
the two-pion correlation function~\cite{hbt}.  Finally, the onset of
the step-like structure in the energy dependence of the inverse slope
parameter of the \mT\ spectra can be reproduced within the
hydrodynamical model by introduction of a rapid change of the
freeze-out conditions at low SPS energies~\cite{ivanov}.  However,
this assumption does not explain the increase of the $T$-parameter
suggested by the RHIC results.  Thus, one can conclude that the models
which do not invoke the onset of deconfinement at the low SPS energies
can not explain the energy dependence of hadron production properties
in central Pb+Pb (Au+Au) collisions.

\section{Summary}

In summary, new results on charged pion and kaon production in central
Pb+Pb collisions at 20$A$ and 30\agev\ were presented and compared to
measurements at lower and higher energies.  A change of energy
dependence is observed around 30\agev\ for the yields of pions and
kaons as well as for the shape of the transverse mass spectra.
Available model explanations are discussed.  At present a reaction
scenario with the onset of deconfinement at low SPS energies best
reproduces the data.
  

\section*{Acknowledgments}

This work was supported by the US Department of Energy
Grant DE-FG03-97ER41020/A000,
the Bundesministerium fur Bildung und Forschung, Germany (06F137),
the Virtual Institute VI-146 of Helmholtz Gemeinschaft, Germany,
the Polish State Committee for Scientific Research (1 P03B 006 30, 1 P03B 097 29, 1 PO3B 121 29, 1 P03B 127 30),
the Hungarian Scientific Research Foundation (T032648, T032293, T043514),
the Hungarian National Science Foundation, OTKA, (F034707),
the Polish-German Foundation, the Korea Science 
\& Engineering Foundation (R01-2005-000-10334-0),
Stichting FOM, the Netherlands,
the Bulgarian National Science Fund (Ph-09/05) and the Croatian Ministry of Science, Education and Sport (Project 098-0982887-2878).




\end{document}